\documentclass[%
 aip,
 amsmath,amssymb,
 reprint,%
floatfix]{revtex4-1}

\usepackage{graphicx}% Include figure files
\usepackage{dcolumn}% Align table columns on decimal point
\usepackage{bm}% bold math
\usepackage[utf8]{inputenc}
\usepackage[T1]{fontenc}
\usepackage{mathptmx}
\usepackage{etoolbox}
\usepackage{siunitx} 
\usepackage{gensymb}
\usepackage{amssymb}
\usepackage{textcomp}
\usepackage{xcolor}
\usepackage{braket}
\usepackage{pdfpages}

\makeatletter
\def\@email#1#2{%
 \endgroup
 \patchcmd{\titleblock@produce}
  {\frontmatter@RRAPformat}
  {\frontmatter@RRAPformat{\produce@RRAP{*#1\href{mailto:#2}{#2}}}\frontmatter@RRAPformat}
  {}{}
\patchcmd{\@outputpage@head}{\@ifx{\LS@rot\@undefined}{}{\LS@rot}}{}{}{}
}%
\makeatother

\begin{document}

%\preprint{AIP/123-QED}

\title{Two-dimensional spin systems in PECVD-grown diamond with tunable density and long coherence for enhanced quantum sensing and simulation}

\author{Lillian B. Hughes}
 \affiliation{Materials Department, University of California, Santa Barbara, Santa Barbara, CA 93117, USA.}

\author{Zhiran Zhang}
 \affiliation{Department of Physics, University of California, Santa Barbara, Santa Barbara, CA 93106, USA.}
 
\author{Chang Jin}
 \affiliation{Department of Physics, University of California, Santa Barbara, Santa Barbara, CA 93106, USA.}

\author{Simon A. Meynell}
 \affiliation{Department of Physics, University of California, Santa Barbara, Santa Barbara, CA 93106, USA.}
 
\author{Bingtian Ye}
 \affiliation{Department of Physics, University of California, Berkeley, Berkeley, CA 94720, USA.}
 
\author{Weijie Wu}
 \affiliation{Department of Physics, University of California, Berkeley, Berkeley, CA 94720, USA.}
 
\author{Zilin Wang}
 \affiliation{Department of Physics, University of California, Berkeley, Berkeley, CA 94720, USA.}
 
\author{Emily J. Davis}
 \affiliation{Department of Physics, University of California, Berkeley, Berkeley, CA 94720, USA.}

\author{Thomas E. Mates}
 \affiliation{Materials Department, University of California, Santa Barbara, Santa Barbara, CA 93117, USA.}

\author{Norman Y. Yao}
 \affiliation{Department of Physics, Harvard University, Cambridge, MA 02138, USA.}

\author{Kunal Mukherjee}
 \affiliation{Department of Materials Science and Engineering, Stanford University, Palo Alto, CA 94305, USA.}

\author{Ania C. Bleszynski Jayich}
 \email{jayicha@ucsb.edu}
 \affiliation{Department of Physics, University of California, Santa Barbara, Santa Barbara, CA 93106, USA.}
\date{\today}

\begin{abstract} 

Systems of spins engineered with tunable density and reduced dimensionality enable a number of advancements in quantum sensing and simulation. 
Defects in diamond, such as nitrogen-vacancy (NV) centers and substitutional nitrogen (P1 centers), are particularly promising solid-state platforms to explore.
However, the ability to controllably create coherent, two-dimensional spin systems and characterize their properties, such as density, depth confinement, and coherence is an outstanding materials challenge. 
We present a refined approach to engineer dense ($\gtrsim$ 1 ppm$\cdot$nm), 2D nitrogen and NV layers in diamond using delta-doping during plasma-enhanced chemical vapor deposition (PECVD) epitaxial growth.
We employ both traditional materials techniques, \textit{e.g.}, secondary ion mass spectrometry (SIMS), alongside NV spin decoherence-based measurements to characterize the density and dimensionality of the P1 and NV layers.
We find P1 densities of 5-10 ppm$\cdot$nm, NV densities between 1 and 3.5 ppm$\cdot$nm tuned via electron irradiation dosage, and depth confinement of the spin layer down to 1.6 nm.
We also observe high (up to 0.74) ratios of NV to P1 centers and reproducibly long NV coherence times, dominated by dipolar interactions with the engineered P1 and NV spin baths.
\end{abstract}

\maketitle
Solid-state spins constitute a powerful platform for quantum technologies.
They can exhibit long quantum coherence, even up to room temperature, are naturally trapped, making them robust and easy to use, and can be simply integrated with sensing targets at nanoscale distances, making them advantageous for quantum sensing.\cite{Degen2017,Schirhagl2014,Rondin2014,Casola2018} 
Further, dense ensembles of coherent spins provide a starting point for investigating strongly interacting spin systems\cite{Rovny2018,Cappellaro2009,Yao2014,Choi2017,Choi2020,Salikhov1981} in which novel, many-body states can arise with applications in both quantum simulation and sensing. 

Two dimensional confinement of dense, coherent spin ensembles opens up a number of experimental avenues. 
Dimensionality plays a critical role in the nature of many-body states, with reduced dimensionality giving access to unique phases and phenomena such as interaction-driven localization.\cite{Yao2014,Yao2018,Chomaz2019,Burin2015,Abanin2019} 
Furthermore, because the 3D angular average of the dipole-dipole interaction yields zero, reduced dimensionality is necessary for collective phenomena such as dipolar-driven spin squeezing\cite{Perlin2020} and plays an important role in the decoherence dynamics of many-body systems.\cite{Davis2021}

For sensing applications, a dense 2D layer of sensors in close proximity to a sensing target exhibits enhanced spatial resolution (set by the depth of the layer) compared to a 3D ensemble at the same volumetric density, while benefiting from either $1/\sqrt{N}$ classical sensitivity enhancements or entanglement-driven enhancements.
Lastly, dense 2D ensembles could serve as a starting point for targeted, on demand formation of individual, optically resolvable defects at specific locations, such as inside nanostructures. 
Altogether, creating thin spin layers with tunable density is of intense current interest but has been minimally explored in solid-state electronic spin systems to date.

In this paper, we explore defects in diamond, an ideal host for solid-state spin qubits.
Diamond's wide bandgap of 5.4 eV, high Debye temperature of 2250 K, and predominantly nuclear spin-free $^{12}$C composition (when grown synthetically using isotopic engineering) provide an exceptionally quiet lattice environment for a myriad of embedded defects with long quantum coherence. 
The most well-explored qubit system in diamond is the negatively charged nitrogen-vacancy (NV) defect, whose spin can be easily manipulated and read out at room temperature using optical and microwave radiation and has been leveraged for a number of notable experiments in quantum sensing.\cite{Degen2017,Schirhagl2014,Casola2018}
Substitutional nitrogen, known as the P1 center, is another diamond defect that may be used for quantum simulation\cite{Davis2021} or building multi-qubit quantum registers \cite{Degen2021} when used in conjunction with the optically addressable NV center.  
While much of the diamond qubit literature focuses on single NV spins for high spatial resolution quantum sensing, dense ensembles of defects (either NV or P1) are becoming increasingly important for a number of experiments leveraging interacting many-body systems.\cite{Zhou2020,Davis2021}
The engineering and characterization of such systems will be the focus of this work. 

The creation of tunably dense, coherent, two-dimensional spin layers in diamond is challenging. 
Ion implantation, the most commonly used method for forming depth-confined defects, suffers from two challenges. 
First, the high energy of the implanted ions results in decoherence-producing collateral damage (\textit{e.g.}, vacancy-related defects),\cite{Tetienne2018} which is exacerbated for the large implantation dosages necessary to achieve high spin densities.
Second, implantation results in a broadened depth distribution of the implanted element. %(as can be seen in Fig.~\ref{fig:growth}b). 
Thin layer confinement can be improved with lower energy implantation, but the spread of depths of these shallowly implanted defects has been observed to be greater than that predicted by SRIM simulations. \cite{Kawai2019, BluvsteinZhang2019} 
More fundamentally, defect properties degrade significantly near the surface,\cite{Myers2014} which is particularly undesirable for quantum simulation where spin-spin interactions must dominate over external decoherence sources. 
Overgrowth after shallow implantation is a path towards improved coherence but risks passivation and loss of defects during the growth.\cite{Findler2020} 

An alternative pathway toward engineering dense, thin spin layers is delta-doping during plasma-enhanced chemical vapor deposition (PECVD) diamond growth. 
This technique, pioneered by Ohno et al.\cite{Ohno2012} for low density ($\lesssim 0.1$ ppm$\cdot$nm) layers and further employed by others,\cite{Bogdanov2021, Jaffe2020, Lobaev2017, Lobaev2018,McLellan2016} allows for control over the dopant thickness and density in a gentle, bottom-up approach that can preserve long coherence. 
Vacancy-related defects can be subsequently promoted via a variety of irradiation methods for further tunability of defect density and location. 
To date, the growth and irradiation approach has predominantly focused on creation of 3D ensembles, \cite{Eichhorn2019,Farfunik2017} with only a few studies investigating dense ($\gtrsim$ 1 ppm$\cdot$nm), thin defect layers where spin-spin interactions may become relevant. \cite{Bogdanov2021,Bogdanov2019,Davis2021} 

In this work, we create highly confined two-dimensional nitrogen layers in diamond via delta-doping during PECVD growth.
Benefiting from an increased understanding of nitrogen incorporation, we identify a set of growth parameters that yields depth confinement of the doped nitrogen layer down to 1.6 nm, as measured by secondary ion mass spectrometry (SIMS). 
We further confirm the 2D nature of the nitrogen spins via spin-based sensing methods, specifically via the decoherence profile of  colocalized NV center probe qubits, which are formed via 200 keV electron irradiation and annealing. 
We next use several methods to quantify the P1 and NV densities and discuss their relative merits. 
From the timescale of the NV qubit decoherence, we measure the P1 density to be 5-10 ppm$\cdot$nm. 
This density measurement is notably 6-8 times smaller than the nitrogen density extracted by SIMS, and we put forth reasoning in support of the accuracy of the qubit-based measurements.
To estimate the NV density, we introduce a new, simple Rabi contrast-based method and benchmark it against other qubit-based density estimation techniques. 
With the Rabi-assisted fluorescence method, we measure an NV density of 1-3.5 ppm$\cdot$nm, tuned via electron irradiation dosage.
We find a high NV/P1 ratio of up to 0.74 and propose a model to support the increased conversion for a 2D nitrogen layer.
Importantly, the NV centers show reproducibly long coherence times, both for ensembles and single NVs, and we find that the coherence is predominantly limited by spin-spin interactions within the dense P1 and NV bath. 
Altogether, this work presents a joint materials and qubit-based approach to engineer and characterize interesting spin systems for applications in quantum technologies.

\section{\label{sec:level2}Material growth, characterization, and preparation} 

\begin{figure}\includegraphics[width=0.45\textwidth]{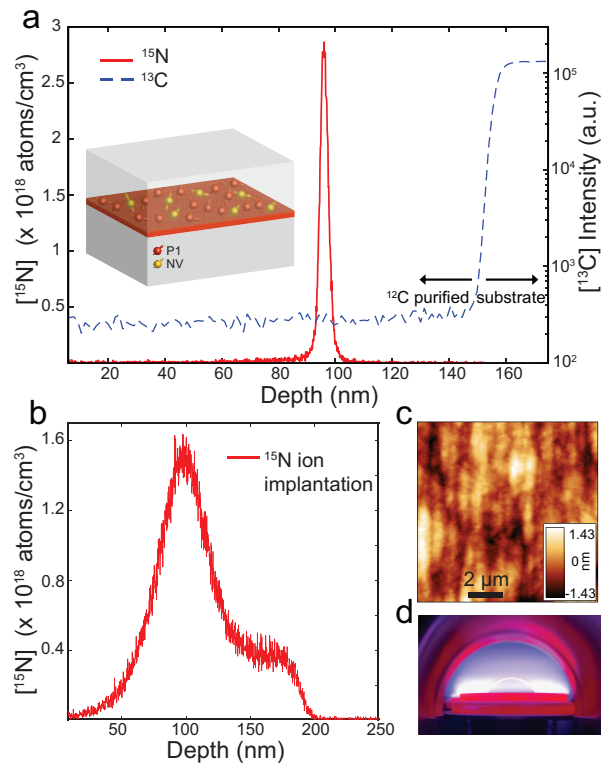}
\caption{\label{fig:growth} (a) SIMS depth profile of PECVD-grown diamond sample A, showing the $^{15}$N (collected as a $^{12}$C$^{15}$N$^-$ ion) and $^{13}$C composition as a function of depth. The FWHM of the dopant layer is 3.55(0.04) nm and the areal density of the peak = 1.057(0.009)$\times10^{12}$ atoms/cm$^{2}$ or 60.1(0.5) ppm$\cdot$nm when integrated over a range of $\pm$3$\sigma$. Inset shows a schematic of a two-dimensional spin system consisting of P1 and NV centers in a delta-doped layer. (b) SIMS depth profile of a $^{15}$N ion implanted sample (90 keV, 1$\times10^{13}$ e$^-$/cm$^{2}$ dose), showing significant broadening compared to the delta-doped sample shown in (a). (c) AFM image of grown sample A, illustrating surface roughness S$_q$ < 0.5 nm over a 100 $\micro$m$^{2}$ region. (d) Image of the PECVD plasma during diamond growth. }
\end{figure}

\subsection*{PECVD diamond growth and nitrogen delta-doping}
Creating nanometer-scale layers and interfaces via in-situ doping during PECVD diamond growth is typically difficult with fast growth rates, on the order of a few ${\micro}$m/hr, and the residual times of the dopant gas.\cite{Achard2020} 
Here we adopt a slow growth approach that allows for precise depth control and promotes high epitaxial quality of the diamond film. 
Through use of a low power (750 W, 25 Torr) plasma, low methane concentration (0.1\% of the 400 sccm H$_2$ flow), and sample holder temperature of $\sim$750\degree~C as measured via a pyrometer, we achieve a slow epitaxial growth rate of 10-30 nm/hr. 
The diamond epilayers are grown on (100) oriented electronic grade diamond substrates (Element Six Ltd.).
Prior to growth, substrates are fine-polished by Syntek Ltd. to a surface roughness of $\sim$200-300 pm and also undergo a 4-5 $\micro$m etch to relieve polishing-induced strain.

During the nitrogen delta-doping period of growth, all plasma conditions are held constant, and $^{15}$N$_2$ gas (1.25\% of the total gas content) is introduced into the chamber briefly to create a delta-doped layer. 
Optical emission spectroscopy (OES) is used to record the residual time of nitrogen in the chamber following doping, which is found to be just 22 seconds.
The $^{15}$N$_2$ doping time in this work is set to be 15 minutes; however, the delta-layer thickness is less than expected given the pre-doping growth rate.
This observation contrasts the assertion that nitrogen doping enhances growth rates\cite{deTheije2000} and suggests that during the delta-doping phase the plasma chemistry changes significantly such that steady growth is slowed or interrupted.
Further investigation is needed to better understand the growth rate discontinuity that allows for thin layer creation and will be the subject of future work.
The $^{15}$N isotope is used to spectroscopically distinguish doped P1 and NV centers from the 99.6\% natural abundance $^{14}$N isotope in the substrate, via the differences in hyperfine coupling to the N nuclear spin. 
We use isotopically purified methane (99.999\% $^{12}$CH$_{4}$, Cambridge Isotope Laboratories) to engineer a diamond lattice environment composed of 99.998\% $^{12}$C, as measured by SIMS (Fig.~\ref{fig:growth}a), thus largely removing $^{13}$C, an isotope with nonzero nuclear spin that contributes magnetic noise and induces NV center decoherence.

A number of parameters are known to affect the density of nitrogen incorporated during growth, including the growth temperature,\cite{Tallaire2015} gas flow ratios,\cite{Lobaev2018} and misorientation angle of the seed substrate.\cite{Meynell2020,Lobaev2018} 
We find that the epitaxial growth rate, which encompasses the effect of all of these parameters, displays the most clear relationship to nitrogen incorporation. 
To control the growth rate, and hence nitrogen incorporation, we modify the miscut angle of the substrate following recent work,\cite{Meynell2020} while keeping temperature and gas flow rates constant. 
We specify the miscut angle by first measuring the virgin substrates using x-ray diffractometry (XRD) rocking curve measurements about the (004) omega peak and subsequently using off-angle fine polishing (Syntek Ltd.) to target a 1.0-1.5\degree~miscut.
In this miscut range, we observe increased nitrogen incorporation with increased growth rate. 

After growth, the diamonds are further electron irradiated and annealed to generate enhanced NV center concentrations. 
Irradiation is performed with the 200 keV electrons of a transmission electron microscope (TEM, ThermoFisher Talos F200X G2 TEM).
This energy is just above the carbon atom displacement threshold energy of 145 keV for (100) diamond,\cite{McLellan2016,Koike1992} thus allowing us to target the formation of single vacancies and minimize extended lattice damage which may occur with higher energies.\cite{Campbell2000} 
The irradiation time is varied to create spots that range in dose from $10^{18}$-$10^{21}$ e$^{-}$/cm$^{2}$. 
The samples then undergo subsequent annealing at $850\degree$~C for 6 hours in an Ar/H$_2$ atmosphere, during which the vacancies diffuse and form  NV centers. 
After irradiation and annealing, the samples are cleaned in a boiling piranha solution (3:1 H$_2$SO$_4$:H$_2$O$_2$) to remove surface contaminants and help stabilize the negative NV$^{-}$ charge state for further measurements. 
All data in the main text of this paper were taken on one sample (sample A). 
To demonstrate repeatability of our growth, preparation, and characterization techniques, we present data on a second, similarly prepared sample (sample B) in the supplemental information (SI). 

\subsection*{Secondary ion mass spectrometry analysis}
The primary bulk materials characterization tool for depth-resolved elemental analysis is SIMS, in which a primary ion beam is directed at a sample, sputtering away material and collecting information about the elemental and isotopic composition as a function of depth. 
We use a CAMECA IMS 7f dynamic SIMS instrument with a primary Cs$^{+}$ beam energy of 7 kV, and current of $\sim$30 nA at an incident angle of $\SI{21.7}{\degree}$ to achieve a sputtering rate of approximately $\SI{20}{\nano\meter/min}$. 
The sample is biased to -3000 V and $^{12}$C$^{15}$N$^{-}$ negative secondary ions are detected using a high mass resolving power, M/$\Delta$M = 6006. 
Only ions from the central $\SI{33}{\micro\meter}$ are collected from the $\SI{100}{\micro\meter}$ sputtering crater to avoid edge effects. 
We observe a delta-doped layer of thickness 3.55(0.04) nm, as determined from the full width at half-maximum (FWHM) of the peak in the nitrogen depth profile illustrated in Fig.~\ref{fig:growth}a. 
The error represents a 95\% confidence interval for a Gaussian peak fit. 
Importantly, the thickness of this layer is significantly narrower than for a sample in which nitrogen was introduced via a standard implantation technique targeting a similar depth, as can be seen in Fig.~\ref{fig:growth}b. 
This sample was implanted with $^{15}$N ions at 90 keV, 1$\times10^{13}$ dose e$^{-}$/cm$^{2}$, and a 7${\degree}$ tilt to mitigate ion channeling. However, channeling is still observed in the SIMS data, as evidenced by the peak's shoulder extending out to 200 nm depth. 
The FWHM of the main peak at 100 nm depth fit with a Gaussian is 53.3(0.2) nm, which is significantly broader than the delta-doped layer, and furthermore broadened compared to the expected 42.2 nm FWHM predicted by SRIM simulation.\cite{2010SRIM} 

SIMS depth resolution is limited by the sample's surface roughness which causes ion mixing at the doped layer interface, and thus the $\SI{3.5}{\nano\meter}$ thickness of sample A is convolved with its surface roughness (Fig.~\ref{fig:growth}c.) and may represent an upper bound. 
However, under similar SIMS measurement parameters we have observed an even thinner layer of 1.61(0.02) nm thickness in a different sample grown with a shorter, five-minute doping time (SI Fig. S1a). 
This observation suggests that the SIMS resolution may not be the dominant limitation in our measurement, but without a secondary characterization, the exact layer thickness remains unknown.

From the SIMS data we can also extract a quantitative estimate of the dopant density by comparing the signal strength to a reference standard with a known dopant dose introduced by ion implantation and calculating a relative sensitivity factor (RSF). 
Taking the background $^{12}$C matrix into account and assuming similar instrumental conditions between runs, the RSF can be used to determine the atomic concentrations of the doped nitrogen. 
Further details on the RSF calibration are given in the SI. 
We integrate over three standard deviations of a Gaussian-fit peak shown in Fig.~\ref{fig:growth}a, to calculate an areal density of 1.057(0.009)$\times10^{12}$ atoms/cm$^{2}$ or 60.1(0.5)  ppm$\cdot$nm. 
Notably, this density should be considered as an estimate only, due to the run-to-run variability in beam tuning, sample charging/surface potentials, and mounting conditions which may cause non-uniform fields.\cite{Wilson1991} 
Additionally, SIMS measures all nitrogen and does not discriminate between different N-related defects, so the extracted density likely overestimates the exact substitutional nitrogen (P1 center) content. 

While SIMS is an important tool for the characterization of delta layers, the shortcomings in estimating doping thickness and concentration requires cross-reference with another measurement. 
However, the sensitivities of conventional techniques such as transmission electron microscopy (TEM) and energy-dispersive x-ray spectroscopy (EDS), or even standard bulk electron paramagnetic resonance (EPR) are generally not low enough to detect the nitrogen densities of delta layers in diamond. 
We note that recent developments in superconducting microcavity based EPR do achieve the required sensitivities,\cite{Ranjan2020} but are still very challenging to implement. 
Furthermore, the insulating nature and hardness of diamond often make ion beam techniques difficult.\cite{Jaffe2020} 
These challenges point toward the need for an alternate characterization route which achieves greater sensitivity in a non-destructive manner, thus leading us to use the NV center itself as a probe of its surrounding lattice.

\section{\label{sec:level3} Characterization of the substitutional nitrogen spin bath}

\begin{figure*}
\includegraphics[scale=0.85]{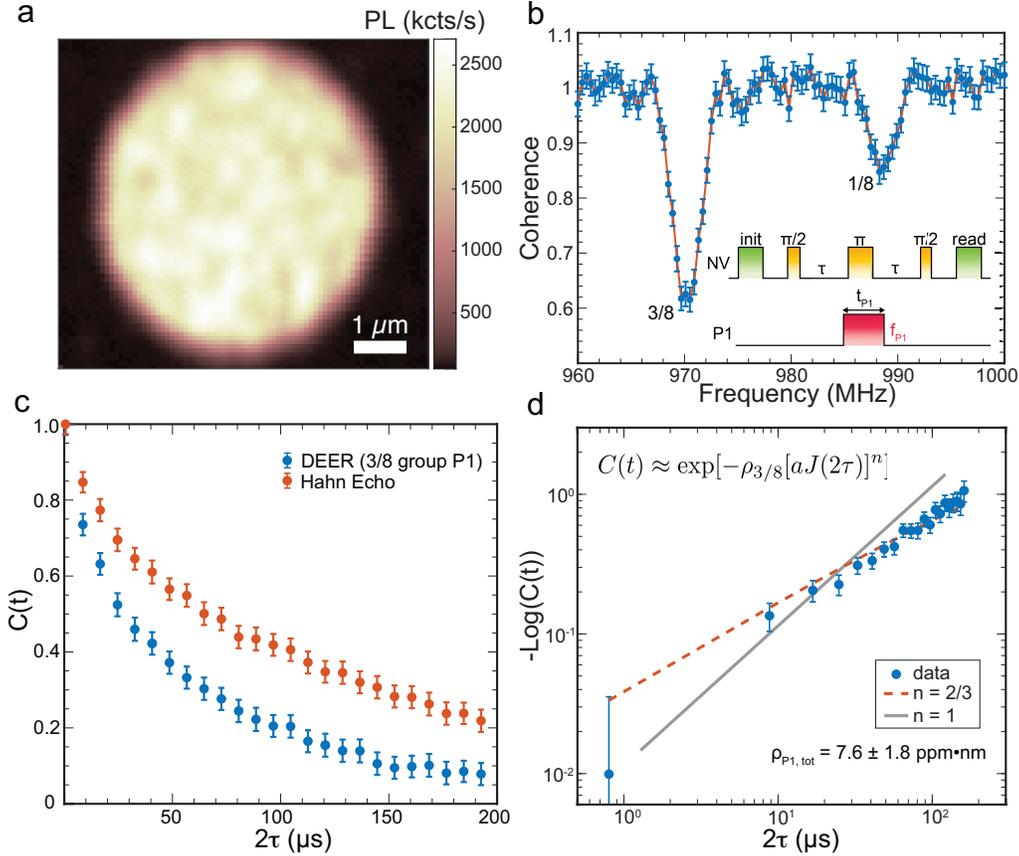}
\caption{\label{fig:P1analysis} (a) Photoluminescence image of a 5-$\micro$m diameter TEM-irradiated spot (5.4$\times10^{19}$ e$^{-}$/cm$^{2}$ dose) on sample A, taken with excitation power of $\sim$1.3 mW (half saturation) measured at the confocal back aperture. This spot is further referred to as spot I in the text. (b) Frequency-swept DEER spectrum of P1 centers at B = 310 G aligned along one of the <111> directions. The frequency range is chosen to zoom in on the ${15}$N $m_I=-1/2$ hyperfine manifold. Inset shows the pulse sequence. The duration of the P1 pulse $t_{P1}$ is set to $\SI{300}{ns}$ while its frequency $f_{P1}$ is swept. The deep (shallow) peak corresponds to the 3/8 (1/8) group of P1 centers. Orange line shows a fit to the data. 
(c) Hahn echo coherence decay of the probe NV centers as a function of total free evolution time 2$\tau$ with (blue) and without (orange) recoupling of the 3/8 P1 group. Error bars represent the standard error of the mean.
(d) Log-log plot of the DEER coherence decay normalized by the Hahn echo data. The orange dashed line shows the best fit for a stretch exponent of $n=2/3$, expected for a 2D bath. For comparison, a best fit for n=1 (a 3D spin bath) is also plotted. The 2D fit with Eq.~\ref{eq1} yields a total P1 areal density $\rho_{P1}$ = 7.6$\pm$1.8 ppm$\cdot$nm. }
\end{figure*}

NV centers serve as excellent sensors of the local spin environment via dipolar interactions with nearby spins. 
In this work, we use the decoherence dynamics of a probe NV center ensemble to reveal characteristics of the surrounding P1 bath. 
We utilize double electron electron resonance (DEER) and Hahn spin echo measurements to accurately quantify the P1 center density, $\rho_{P1}$,\cite{Eichhorn2019,Osterkamp2020,Davis2021} and confirm their dimensionality.\cite{Davis2021} 
Two-dimensionality is defined by a layer thickness that is smaller than the inter-spin spacing for a given dopant density.
Fig.~\ref{fig:P1analysis}a shows a photoluminesence image of the TEM-irradiated, NV-rich spot on sample A (henceforth referred to as spot I). 
All NV-based measurements are performed on a home-built confocal microscope using a 532-nm diode laser and an external magnetic field of 310 G aligned along one of the NV <111> axes. 
Radiofrequency (RF) signals are delivered through a gold antenna fabricated on a glass coverslip and placed underneath the diamond.  

\subsection*{Extraction of density and dimensionality}
Through DEER-based sensing, we characterize the nitrogen spin bath by selectively probing the quasistatic contribution of P1 centers to NV decoherence. 
Fig.~\ref{fig:P1analysis}b shows a frequency-swept DEER spectrum of the $m_I=-\frac{1}{2}$ hyperfine state of the $^{15}$N P1 centers\cite{Eichhorn2019} revealing two peaks: the shallow peak (the 1/8 group) corresponds to the $^{15}$N P1 center with its Jahn-Teller axis aligned to the external magnetic field and the deeper peak (the 3/8 group) corresponds to the three other degenerate $^{15}$N P1 centers with Jahn-Teller axes misaligned to the field.\cite{Davies1981} The DEER pulse sequence is illustrated in the inset of Fig.~\ref{fig:P1analysis}b. 
A pump pulse resonant with the 3/8 group of P1 centers with frequency f$_{P1}$ recouples the P1 bath to the NV centers, and the resulting contribution to decoherence is read out via an NV center Hahn echo measurement. 
Essentially this sequence constitutes a Ramsey measurement of the NV-P1 interaction. 

The timescale of the DEER decoherence decay provides information about the density of P1 centers, and the stretch exponent of the decay indicates the dimensionality of the spin bath. 
According to theory presented in Ref.\cite{Davis2021} the decoherence decay profile is expected to take the following form:
\begin{equation}\label{eq1}
    C(t)\approx\exp[-\rho_{3/8} [aJ(2\tau)]^{n}]
\end{equation}
where $\rho_{3/8}$ is the density (in units of ppm$\cdot$nm for a 2D spin bath) of the driven group of P1 centers, $a$ is a dimensionless constant (that equals 2.626 for a 2D spin bath or 3.318 for a 3D spin bath), and $J=2\pi\times52\text{ MHz·nm}^3$ is the dipolar coupling strength. 
The stretch power, $n=d/\alpha$, depends on the dimensionality of the system, $d$, and the power law of the interactions, $\alpha$. 
Importantly, Eq.~\ref{eq1} is only valid at times short compared to the correlation time $\tau_c$ of the bath, \textit{i.e.} $\tau\ll\tau_c$. 
For a 2D spin bath coupled to the NV via dipolar interactions ($\alpha=3$), we expect $n=2/3$ when measuring on a time scale much shorter than $\tau_c$; for a 3D bath we expect $n=1$ as seen, \textit{e.g.}, in Ref.\cite{Davis2021}.
Further details on the theoretical analysis are given in the SI.

Fig.~\ref{fig:P1analysis}c plots the NV coherence as measured by Hahn echo and by DEER using a differential measurement scheme.\cite{Dolev2019} 
The DEER data shows the expected reduction of coherence when recoupling the P1 centers. 
Fig.~\ref{fig:P1analysis}d plots the DEER decay normalized by the NV Hahn echo on a log-log scale as a function of free evolution time ($2\tau$). 
The DEER data is normalized by the Hahn echo in order to eliminate any contributions to decoherence from other sources.
The excellent agreement of the data to a fit with stretch power of $n=2/3$ and poor agreement with $n=1$ confirms the two-dimensional nature of the spins. 

The data also provide a quantitative estimate of the total P1 density, $\rho_{P1} = 7.6\pm1.8$ ppm$\cdot$nm in the TEM irradiation spot I.
This density corresponds to a P1 spacing of $\approx$13.8 nm, larger than the layer thickness as measured by SIMS and hence consistent with a 2D spin system.
Density measurements are averaged over six locations within TEM spot I, where the 1.8 ppm$\cdot$nm standard deviation reflects the $\rho_{P1}$ variation across the spot.  
We attribute this variation to inhomogeneous nitrogen incorporation during growth, which will be a subject of future study. 
We note that the NV-DEER based estimate of $\rho_{P1}$ is $\sim 8$ times lower than the SIMS measurement (60.1(0.5) ppm$\cdot$nm). 
This trend holds for the data collected on sample B as well, where the DEER-measured $\rho_{P1}$ of 16 ppm$\cdot$nm is $\sim 6$ times less than the SIMS-estimated 89.5(0.8) ppm$\cdot$nm (full analysis in the SI). 
Given the uncertainties in the RSF value and the fact that SIMS measures all forms of nitrogen and is not specifically sensitive to P1 centers (as in DEER), this discrepancy is not surprising, and importantly, it confirms our understanding that SIMS needs a cross-referencing technique to more accurately determine the concentration of dopant atoms in our material. 

\section{\label{sec:level4}Characterization of NV center properties}
We next move to characterizing the density, dimensionality, and coherence of NV centers formed via our growth and irradiation approach. 
In addition to enabling the study of strongly interacting, optically addressable spin ensembles in 2D for sensing and simulation, it also facilitates the creation of coherent NV centers with a specified, tunable density to realize \textit{e.g.}, a single NV per nanopillar in a scanning probe tip\cite{Maletinsky2012,Pelliccione2016} or a single NV inside of a photonic crystal cavity.\cite{Englund2010}
Here we explore the creation and properties of the NV centers in our delta-doped system, finding both high NV/P1 ratios and remarkably long coherence which is predominantly limited by the P1 and NV environment.

\begin{figure}
\includegraphics[scale=0.95]{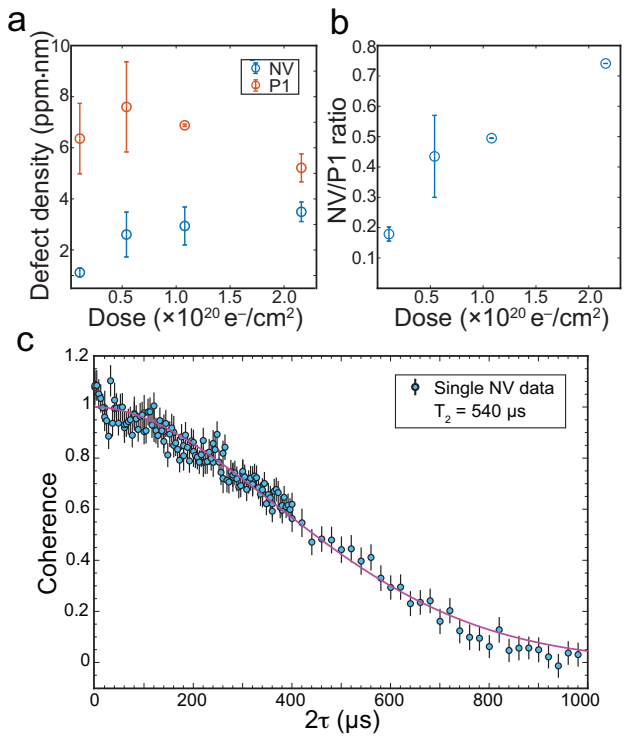}
\caption{\label{fig:NVanalysis} (a) Plot of the $\rho_{P1}$ and $\rho_{NV}$ (measured with DEER and Rabi-assisted fluorescence counting, respectively) as a function of electron irradiation dose. Error bars show the standard error of the mean. (b) NV/P1 ratio as a function of electron irradiation dose. Small standard error bars are due to a small sample size. (c) Hahn echo decay for a single NV center outside of the TEM spots illustrating a T$_{2}$ of 540 $\micro$s.}
\end{figure}

\subsection*{NV center density estimation}
A number of methods are typically used to estimate the NV center density, $\rho_{NV}$, when individual NV centers are not optically resolvable and countable. 
Broadly, these methods fall into two categories: the first involves probing the decoherence timescale of the NV ensemble under a specific RF pulse sequence, such as NV-NV DEER ,\cite{Eichhorn2019} instantaneous diffusion,\cite{Eichhorn2019} or an XY-8 sequence.\cite{Zhou2020} 
The second, all-optical method estimates density from the ensemble fluorescence intensity. 
However, these techniques all come with their drawbacks. 
The spin decoherence methods require long averaging times and are limited to dense NV ensembles where NV-NV dipolar interactions are comparable to or stronger than other sources of decoherence.
Biased estimates can result when ensembles deviate from this high (NV-dominant) density, and pulse errors can introduce further bias (see SI for further explanation and the results of implementing these methods on samples A and B).
Estimating density of NV ensembles via fluorescence can be hampered by unwanted background fluorescence from other irradiation-induced defects, which becomes particularly relevant at high electron dosages.

Here we introduce a simple and more reliable approach to determine the NV density. 
Our method benefits from not only its quick and easy implementation but also its large dynamic range which allows for density estimation from single NVs to high-density ensembles.
We measure the fluorescence intensity difference between NV ensemble spin states ($m_s=0$ and $m_s=\pm1$) in a Rabi experiment and normalize by the corresponding signal from a single NV. 
By isolating the component of the fluorescence that changes with microwave driving, the measurement is immune to contributions from non-NV$^{-}$ defects.
This method combines fluorescence counting and spin manipulation to measure a uniquely NV signal. 

Briefly, our NV density estimation technique proceeds as follows (further details provided in the SI).
First the point spread function (PSF) of the confocal setup is determined by fitting the fluorescence profile of an average of several single NV centers. 
Next, the PSF is convolved with a random distribution of 1 ppm$\cdot$nm NVs to calculate a spatial average of the resulting signal, $f_{avg}$.
This average factor accounts for the fact that not all NV centers in the confocal spot reside at the center of the spot. 
Lastly $\rho_{NV}$ is estimated by relating $\Delta PL_R$, the fluorescence intensity difference of the $m_s=0$ and $\pm1$ states in a Rabi experiment, of single and ensemble NVs via:
\begin{equation}
    \rho_{NV}= \frac{\Delta PL_{R, NV ensemble}}{\Delta PL_{R, single NV}}\times\frac{1}{f_{avg}}\times4
\end{equation}
The factor of four accounts for the four possible NV orientations, assuming an equal distribution among them, to yield $\rho_{NV}$ as the total NV density. 

Applying this method to the TEM spot I, we find $\rho_{NV}=2.6\pm0.9$ ppm$\cdot$nm averaged over eight different laser positions within the spot. 
The variation of $\rho_{NV}$ across the spot is consistent with the variation observed in $\rho_{P1}$.
In the SI, we compare measurements of $\rho_{NV}$ in samples A and B obtained with our Rabi-assisted fluorescence method and spin decoherence methods, specifically NV-NV DEER and XY-8. 
Using our new method, we find $\rho_{NV}$ to lie between the estimates produced by NV-NV DEER and XY-8 measurements; this trend is consistent with our expectations for the biases imposed by these two methods and advocates for the accuracy of our Rabi-assisted fluorescence measurement. 

\subsection*{NV center creation and coherence}
We next characterize $\rho_{NV}$ across a range of electron irradiation dosages and importantly find a high NV/P1 ratio of up to 0.74, corresponding to $\rho_{NV}$ of up to 3.5$\pm$0.4 ppm$\cdot$nm.
Figs.~\ref{fig:NVanalysis}a-b show a clear increasing trend of $\rho_{NV}$ and NV/P1 ratio with electron dosage. 
The NV/P1 ratio is an important figure of merit for this work because the NV and P1 centers are the dominant contributors to decoherence in our samples, as will be discussed below. Further, $\rho_{NV}$ and $\rho_{P1}$ are quantitatively probed by our NV-based measurements.
Interestingly, we observe higher NV/P1 ratios in our 2D delta-doped sample compared to 3D nitrogen-doped samples prepared via similar irradiation dosages and growth procedures (with longer doping times).\cite{Eichhorn2019} 
We posit that the higher NV/P1 ratios observed here are related to the dimensionality of the nitrogen layer and the diffusion of vacancies during annealing. 
In a 2D layer, each nitrogen interacts with a greater number of vacancies compared to the 3D case because of the finite diffusion volume during annealing. 
Thus, the enhancement will be determined by a dimensionless parameter given by $V_D/t_L$, where $V_D$ is the diffusion volume and $t_L$ is the layer thickness. 

To test the validity of this explanation, we construct a simple vacancy capture model (detailed in the SI) where vacancies migrate according to a weighted one-dimensional random walk and can be captured by an adjacent nitrogen with unit probability. 
Using our experimental parameters, we find a 15$\times$ enhancement of conversion for the 2D layer vs. the 3D layer, supporting the hypothesis that the geometrical diffusion effect is responsible for the observed enhancement in 2D.  
While this model does not attempt to rigorously simulate all aspects the NV center formation process, it supports our experimental findings of increased conversion in 2D-doped samples and suggests an interesting subject for future study.

Beyond the high NV/P1 ratios, a notable feature of the NV centers in our delta-doped samples is their long Hahn echo T$_2$ coherence time across a range of P1 and NV defect densities (see SI Fig.~S5). 
Maintaining long coherence is crucial for all the aforementioned experiments and applications in quantum sensing and simulation, particularly when operating in the dense P1 or NV regime.
In the unirradiated regions, we find single, optically resolvable delta-doped NV centers in a dense P1 background, and we measure Hahn echo T$_2$ times of 535 $\micro$s, 562 $\micro$s, and 489 $\micro$s for three single NVs, a representative example of which is shown in  Fig.~\ref{fig:NVanalysis}c. 
This result is promising for applications such as NV scanning probe microscopy, which demands single-NV-containing probes formed with high yield. 
Higher density ensembles of NV centers in TEM spot I show an average coherence time of 108$\pm$10 $\micro$s, which is in good agreement with the expected decoherence from a bath of P1 and NV centers of the measured density (see SI for analysis of the P1 and NV contributions to decoherence).
The dominance of intentional spin-spin interactions, combined with the system's 2D nature, is critical for experiments in many-body sensing and simulation.

\section{\label{sec:level5} Conclusion}
In this work, we have demonstrated the controlled creation and characterization of 2D spin systems in diamond with tunable density and long coherence.  
By harnessing slow growth rates and substrate miscut control, we achieve thin nitrogen layer confinement and characterize the P1 dimensionality and density with both SIMS and NV spin decoherence-based techniques, noting the importance of spin-based measurements in characterizing the relevant spin bath density. 
We demonstrate NV density tunable via electron irradiation and observe high NV/P1 ratios in the 2D system.
NV density characterization is performed using a simple Rabi-assisted, fluorescence-based NV method introduced here that does not suffer from the biases of other estimation techniques and demonstrates a large dynamic range.
Furthermore, we observe long NV T$_2$ coherence times despite the rich P1 and NV environment, suggesting that the system is not severely limited by disorder from other defects.
Altogether this work presents a careful materials and qubit-joint approach to engineer unique spin systems for enabling a variety of investigations in quantum sensing and simulation.

\section*{Supplementary Information}
See the SI for additional details on the SIMS analysis, characterization of sample B, details of the Rabi-assisted density estimation and comparison to other methods, vacancy diffusion/NV creation model, additional coherence measurements and interpretation, and supporting information about the decoherence decay analysis.

\section*{Data Availability}
The data that support the findings of this study are available from the corresponding author upon reasonable request.

\begin{acknowledgments}
We gratefully acknowledge support of the U.S. Department of Energy BES grant No. DE-SC0019241 (materials growth and characterization), the Center for Novel Pathways to Quantum Coherence in Materials, an Energy Frontier Research Center funded by the U.S. Department of Energy, Office of Science, Basic Energy Sciences (coherence studies), and the Army Research Office through the MURI program grant number W911NF-20-1-0136 (theoretical studies).
We acknowledge the use of shared facilities of the UCSB Quantum Foundry through Q-AMASE-i program (NSF DMR-1906325), the UCSB MRSEC (NSF DMR 1720256), and the Quantum Structures Facility within the UCSB California NanoSystems Institute.
L. B. H. acknowledges support from the NSF Graduate Research Fellowship Program (DGE 2139319) and the UCSB Quantum Foundry.
S. A. M. acknowledges support from Canada NSERC (AID 516704-2018) and the UCSB Quantum Foundry. 
E. J. D. acknowledges support from the Miller Institute for Basic Research in Science.
\end{acknowledgments}

\bibliography{2D_revised}

\end{document}

\doublespacing
\maketitle

\section{Additional SIMS details}

\subsection{Depth resolution}
All of the SIMS data presented in the main text was taken using a 7 kV (30 nA) Cs$^{+}$ ion beam (sputter rate $\approx\SI{20}{\nano\meter/min}$). 
We found that this lower energy beam, compared to a more standard 15 kV, allows for improved resolution of the delta-layer thickness. 
This is because of reduced ion mixing that occurs with a smaller impact energy. 
The disadvantages of using lower voltage are decreased sensitivity (smaller secondary ion signal), slower etch rates, and broader beam size, but none of these critically impact our analysis. 

\begin{figure}[ht]
    \centering
    \includegraphics{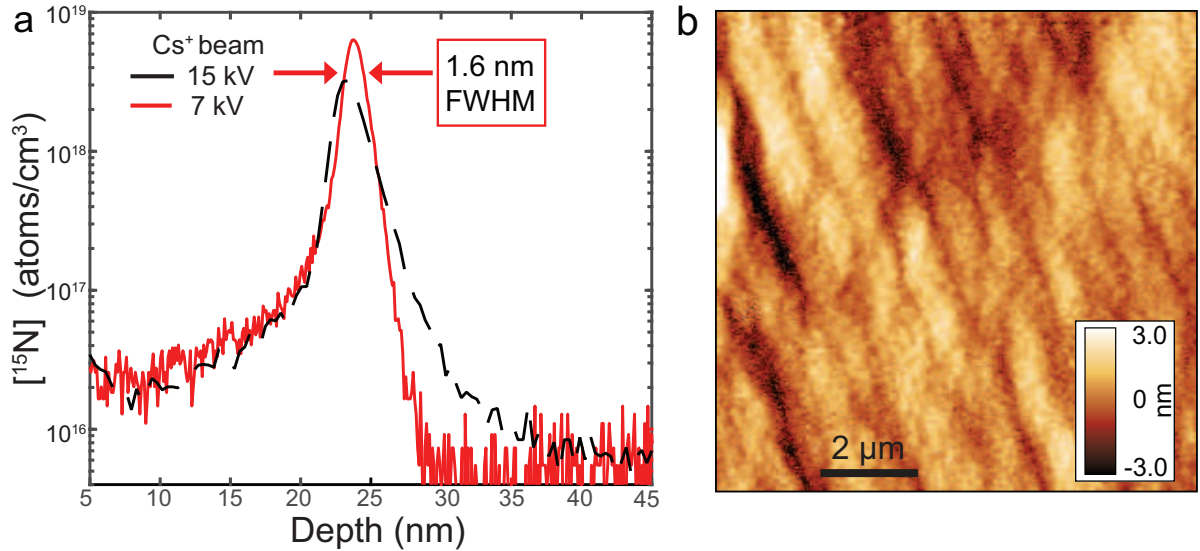}
    \caption{(a) SIMS $^{15}$N depth profile showing a $^{15}$N delta-doped layer with FWHM thickness of 1.61(0.02) nm. The resolution is improved from 2.42(0.07) nm by using a lower energy Cs$^{+}$ ion beam (red shows 7 kV and black dashed shows 15 kV). (b) AFM of the sample in (a) showing surface roughness S$_q$ $\approx$ 821 pm over a 100 $\mu$m$^{2}$ region.}
    \label{fig:S1}
\end{figure}

Fig.~S\ref{fig:S1}a shows 1.6 nm-thick layer that previously was broadened to 2.4 nm using the 15kV beam. 
This particular sample was grown with a shorter nitrogen doping time (5 min) than the primary main text sample A (15 min), which contributes to the thinness of the layer. 
The surface roughness of the two samples are comparable, which suggests that the surface may not be limiting the depth resolution significantly beyond 1.6 nm.

\subsection{RSF calculation}
To estimate nitrogen concentration with SIMS, we employ a relative sensitivity factor (RSF) which is a function of the secondary ion intensity observed for the species of interest and the surrounding matrix environment, as described by the following expression
\begin{equation}
    RSF = (I_M/I_D)*C_D/a_M
\end{equation}
 where I$_M$ and I$_D$ indicate intensity of the matrix and dopant, respectively, C$_D$ is the dopant concentration, and $a_M$ is the abundance of the reference matrix isotope.\cite{Wilson1991}
Comparing to a reference standard, the RSF can be calculated according to the equation:
\begin{equation}
    RSF = \frac{\phi EM I_M}{r a_M FC \sum{I_D}dz}
\end{equation}
where $\phi$ indicates the implantation fluence, r the sputter rate, $\sum{I_D}dz$ is the integral of the implanted depth profile, and EM/FC are electronics factors related to the electron multiplier and Faraday cup detector systems.\cite{Wilson1991} 
Using an implanted sample with a nitrogen dose of 1$\times10^{13}$ atoms/cm$^2$ and the 7 kV Cs$^{+}$ beam, we determine an RSF of 1.375$\times10^{22}$ for the $^{12}$C$^{15}$N$^{-}$ ion referenced to a natural abundance $^{12}$C matrix. 
Wilson et al. report uncertainties in RSF values of $\pm$20-30\% even under controlled experimental conditions, suggesting that altogether dopant quantifications in SIMS are best interpreted as an estimate only.\cite{Wilson1991} %note that Wilson reports 2.5e22 for a 14.5 kV Cs+ beam, which is similar but obviously there is some disparity

\section{Sample B characterization}
To demonstrate repeatability of our growth, preparation, and characterization techniques, we prepared a twin sample for additional analysis. 
Sample B was grown with the same growth conditions as sample A (750 W, 25 torr, 750\degree C plasma, 0.4$\%$ methane addition, and 1.25$\%$ $^{15}$N$_2$ doping for 15 minutes). 
It was similarly irradiated with 200 keV electrons in a TEM and annealed at $850\degree$C for 6 hours in an Ar/H$_2$ atmosphere. 

\begin{figure}
    \centering
    \includegraphics{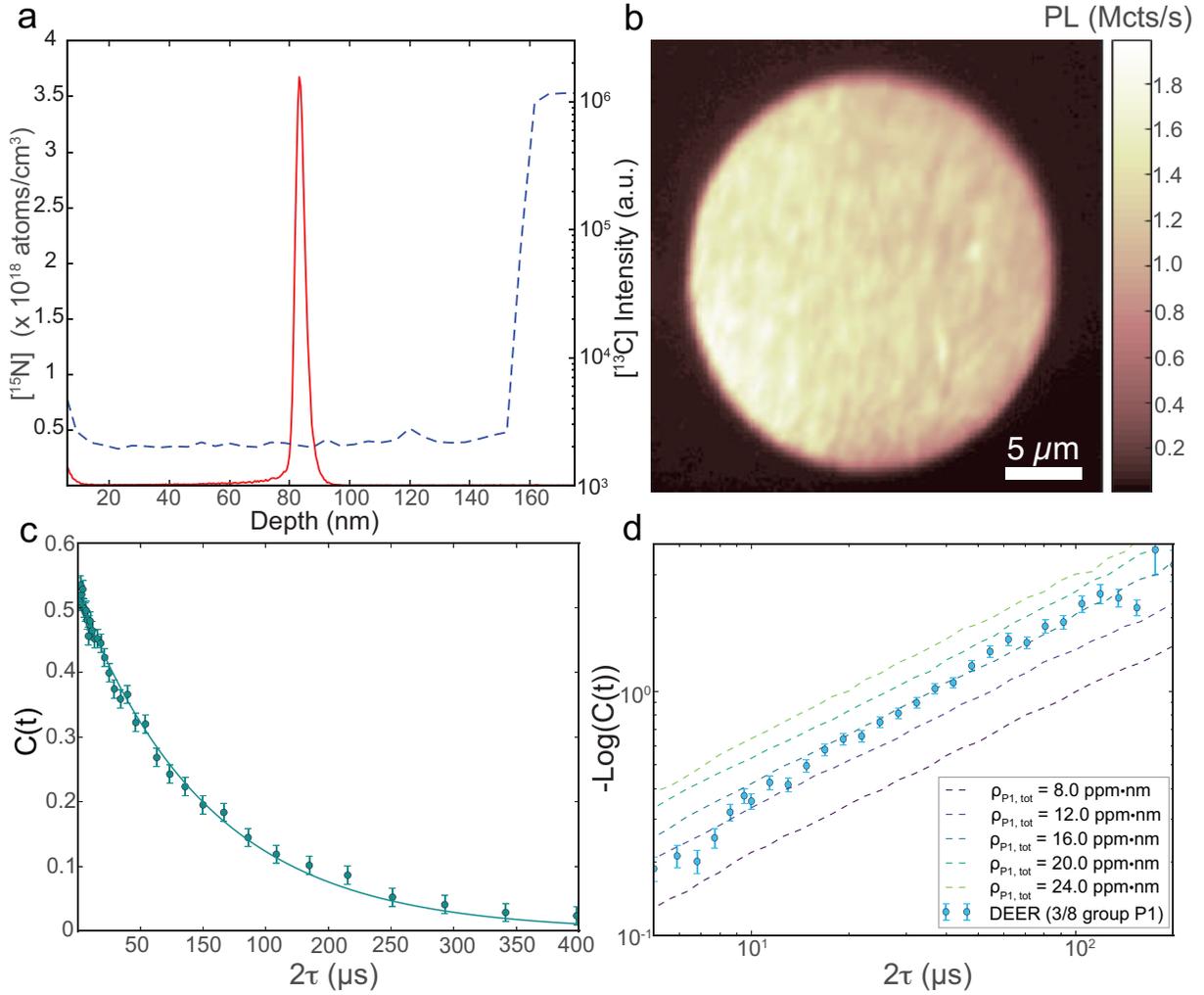}
    \caption{(a) SIMS depth profile showing the $^{15}$N and $^{13}$C composition as a function of depth in the diamond PECVD epitaxy of sample B. 
    The thickness of the dopant layer FWHM is 4.06(0.07) nm when measured with a 15 kV ion beam, and the areal density of the peak = 6.2$\times10^{12}$ at/cm$^{2}$ or 89.5(0.8) ppm$\cdot$nm when integrated over a range of $\pm$3$\sigma$. 
    (b) Scanning confocal photoluminescence image of a TEM-irradiated spot on sample B. 
    (c) Representative Hahn echo decay of an NV center in the TEM spot, showing a T$_{2}$ of 103$\pm$2 $\micro$s.  
    (d) Log-log plot of the P1 DEER coherence decay. The dashed lines show fits of varying P1 areal densities where the stretched exponent is fixed to be $n=2/3$, as expected for a two-dimensional bath. 
    The best fit line yields a total P1 areal density $\rho_{P1}$ = 16.0 ppm$\cdot$nm.}
    \label{fig:S2}
\end{figure}

Fig.~S\ref{fig:S2}a. shows SIMS characterization of sample B, where the delta-doped layer FWHM is 4.06(0.07) and the areal density (calculated by integrating over a range of $\pm$3$\sigma$ from the peak center) is 89.5(0.8) ppm$\cdot$nm. 
This data was collected with a 15 kV energy beam, as it was taken before realizing the resolution benefit of 7 kV, and thus the FWHM thickness represents an upper bound on the true layer thickness. 
Fig.~S\ref{fig:S2}b shows a TEM-irradiated spot (6.8$\times10^{19} e^{-}$/cm$^{2}$) used for further characterization of the defect and spin bath in sample B. 
Fig.~S\ref{fig:S2}c shows a typical T$_2$ coherence time of 103$\pm$2 $\micro$s for the NV ensemble in the TEM spot, and Fig.~S\ref{fig:S2}d contains a log-log plot of the DEER coherence decay. 
The dashed lines show fits of varying P1 areal densities where the stretched exponent is fixed to be $n=2/3$, as expected for a two-dimensional bath. 
The best fit line yields a total P1 areal density $\rho_{P1}$ = 16.0 ppm$\cdot$nm.
Characterization of the NVs in sample B is discussed in the NV density estimation methods section of this SI.

\section{NV density estimation using the Rabi-assisted fluorescence method}
\begin{figure}
    \centering
    \includegraphics[scale=0.95]{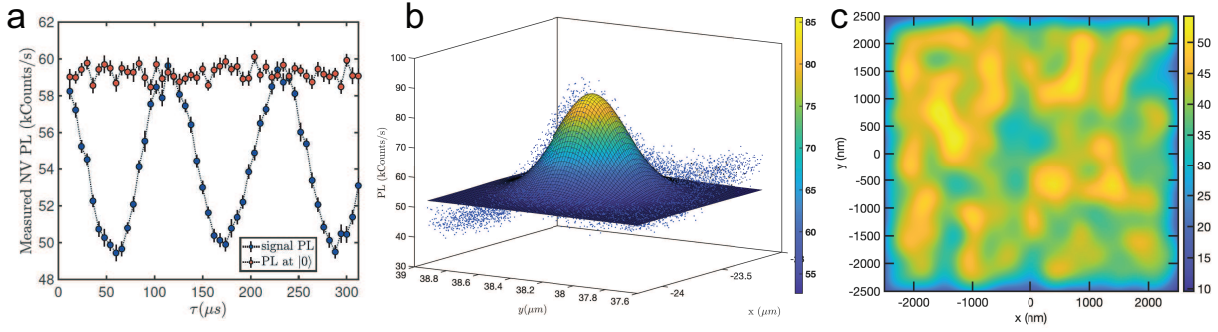}
    \caption{(a) Rabi measurement on a single NV center. (b) Fluorescence profile from a single NV center used to determine the confocal PSF. (c) Simulated fluorescence of a 1ppm$\cdot$nm NV ensemble that is convolved with the single NV PSF.
    }
    \label{fig:S3}
\end{figure}

As described in the main text, our method for estimating the NV density relies on measuring the fluorescence intensity difference of the $m_s=0$ and $m_s=\pm1$ states ($\Delta PL_R$) for a single NV center during a Rabi experiment and extrapolating this to an ensemble of NVs. 
We measure four single NVs of the same orientation and find $\Delta PL_R$ = 8.7(4) kCounts/s (example shown in Fig.~S\ref{fig:S3}a). 
Since the photoluminescence of both the single and ensemble NV centers exhibits saturation behavior relative to the excitation beam intensity, we set the laser power to be at half the saturation power for consistency across all of our measurements.

Next, we determine the relevant point spread function (PSF) for our confocal setup. 
In an ideal situation where the resolution of the confocal image is diffraction limited, the PSF of an atomic defect is best modeled by the Airy disk function. 
However, to account for aberrations caused by various non-ideal optical components, we use the 2D Gaussian function to model our PSF:
\begin{equation}\label{eq:1}
    PSF(x,y) = A * exp[-\frac{(x-x_0)^2}{r_1^2}]*exp[-\frac{(y-y_0)^2}{r_2^2}]+c,
\end{equation}
where A is the amplitude of the Gaussian beam, $r_1$ and $r_2$ are the Gaussian radius characteristic of the beam, x$_0$ and y$_0$ are determined by the location of the NV, and c accounts for the photoluminescence of the background. 
By fitting the single NV confocal image shown in Fig.~S\ref{fig:S3}b. with Eq.\ref{eq:1}, we can obtain the desired model of the setup’s PSF.

Because not all NV centers reside at the center of the confocal spot, and the polarization of the incident light affects the observed brightness of different NV orientations, it important to take the single NV PSF into account when measuring an NV ensemble.
The PSF is then convolved with a random distribution of 1 ppm$\cdot$nm NVs, which results in the simulated confocal image shown in Fig.~S\ref{fig:S3}c.
A spatial average of the resulting image is taken to determine the conversion factor,  $f_{avg}$ = $\SI{38(8)}{ppm^{-1}nm^{-1}}$.
This conversion factor then allows us to relate the measured $\Delta PL_R$ for an NV ensemble to a single NV to estimate the ensemble density, according to Eq. 2 in the main text.

\section{Additional NV density estimation methods}
As discussed in the main text, there are a number of approaches to estimate the NV density, $\rho_{NV}$. Here we describe in more detail the spin coherence-based methods of XY-8 and NV-NV DEER and their inherent biases. 

%(Edited by Weijie)
In an (XY-8)n measurement, the NV group aligned to the external magnetic field is treated as an ensemble of both probe and bath spins. 
Ideally, the XY-8 sequence perfectly decouples interactions between this target NV group and any remaining spins in the system; thus, the XY-8 decoherence is induced by the interactions only between probe spins, whose density can therefore be estimated via the $T_2^{XY8}$ timescale. 
In practice, the interactions between the probe spins and other spins are not perfectly cancelled by the XY-8 pulse sequence. 
Our model approximates the decoherence as arising purely via probe spin interactions, and hence this method systematically overestimates the NV density. 
This overestimate is evident in the data for both samples A and B, where the XY-8 measured NV densities are large compared to other methods (see Table~\ref{tab:samplecompare}).

\begin{figure}
    \centering
    \includegraphics[scale=0.95]{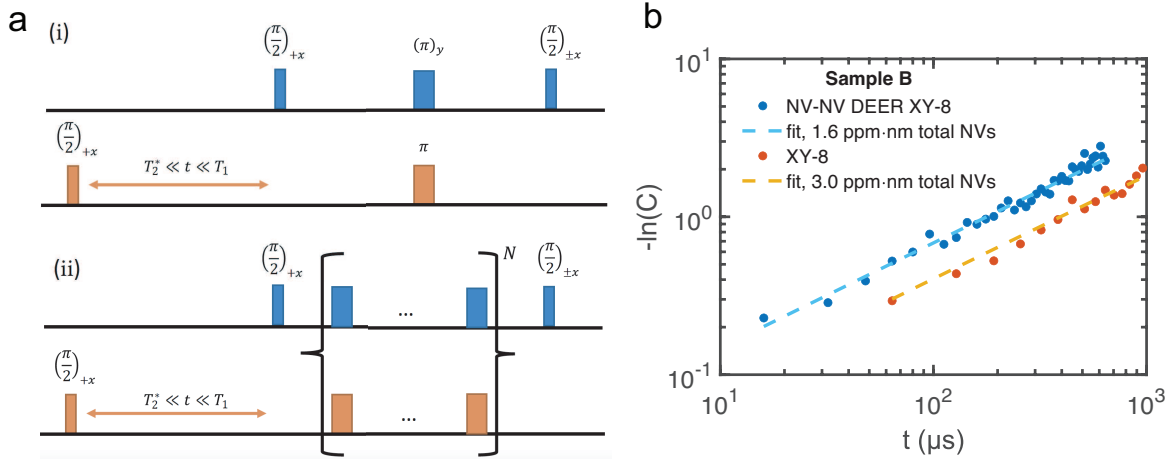}
    \caption{(a) Pulse sequences for NV-NV DEER (i) and NV-NV DEER XY8 (ii). (b) Estimation of $\rho_{NV}$ in sample B using NV-NV DEER XY-8 (blue) and XY-8 (orange).
    }
    \label{fig:S4}
\end{figure}

In both the NV-NV DEER and NV-NV DEER XY-8 experiments, the NV group aligned to the external magnetic field acts as a probe ensemble for the three degenerate NV groups. 
After laser polarization, all four NV groups are initialized into the ground state $\ket{0}$, which is followed by a $\pi/2$ pulse driving the three degenerate bath NV group into the superposition state $(\ket{0}+\ket{-1})/\sqrt{2}$. 
We then wait for a time $T_2^*\ll t\ll T_1$ such that the bath NVs fully decohere while the probe ensemble remains polarized. 
After initialization, NV-NV DEER is performed via a Hahn echo sequence on the probe spins with an additional $\pi$ pulse on the bath NVs.
These pulse sequences are illustrated in Fig.~S\ref{fig:S4}a.

\begin{table}[]
    \centering
    \begin{tabular}{|c|c|c|c|}
    \hline
     & Rabi-assisted (ppm$\cdot$nm) & XY-8 (ppm$\cdot$nm) & NV-NV DEER (ppm$\cdot$nm)\\
     \hline\hline
    sample A & 2.6(9) & 4.0(2) & 1.4(2) \\
    \hline
    sample B & 3.6(2) & 3.0(2) & 1.6(1) \\
    \hline
    \end{tabular}
    \caption{Comparison of NV density estimation methods. Data were taken on similar dose TEM spots (5.4$\times10^{19}$ e$^{-}$/cm$^{2}$ for sample A and 6.8$\times10^{19}$ e$^{-}$/cm$^{2}$ for sample B). The data show a trend toward overestimation using the XY-8 method and underestimation with NV-NV DEER.}
    \label{tab:samplecompare}
\end{table}

Alternatively, we may perform NV-NV DEER XY8 after initialization. 
We apply identical XY-8 pulses to both the probe and the bath NVs, such that the probe-bath NV-NV Ising interaction survives while interactions between the probe and, e.g., dark P1 spin ensembles are decoupled. 
Compared to NV-NV DEER, the NV-NV DEER XY-8 sequence decouples these latter interactions more effectively, such that  the decoherence is dominated by the probe-bath NV-NV Ising interaction. 
In both pulse sequences, the probe-probe NV interactions and the probe-bath NV interactions all contribute to the $T_2$ decay signal, \textit{i.e.} the decoherence timescale reflects the total density of probe and bath NVs. 
 
Additionally, the projections of microwave polarization on the three NV orientation groups are different, so the Rabi frequencies of the three NV groups are not identical. 
Therefore, $\pi$ pulses cannot be simultaneously calibrated for all of the three groups. 
When the $\pi$ pulse for the bath spins is not perfect, the Ising interaction between the probe and bath spins are not fully recoupled and the probe bath is expected to have a longer decoherence time scale. 
As a result, the NV-NV DEER methods will lead to an underestimation of the defect density (as can be seen in Table \ref{tab:samplecompare}). 

As shown in Fig.~S\ref{fig:S4}b for sample B, the NV-NV DEER XY-8 measurement gives an estimation for $\rho_{NV}$ of 1.6(1) ppm$\cdot$nm, which should be a lower bound. 
The (XY8)n measurement gives 3.0(2) ppm$\cdot$nm, which should be an upper bound. 
%The Rabi-assisted fluorescence method yields a result of 3.6(2) ppm$\cdot$nm. 
We also performed and compared these three methods in sample A, and we find a similar result (raw data not shown but reported in Table~\ref{tab:samplecompare}). 
Altogether, we find that the Rabi-assisted method is unaffected by the biases of the spin coherence-based techniques and thus provides a quick estimate of the NV density in our samples.

\section{2D versus 3D conversion simulation details}
To explore the enhanced P1-to-NV conversion that we observe experimentally in the 2D delta-doped samples relative to 3D-doped samples, we construct a simple vacancy capture model where vacancies migrate according to a one-dimensional random walk.
We initialize vacancies randomly throughout a 10 $\micro$m depth.
Then, for each vacancy, we perform a random walk in z for 1.24$\times10^{8}$ steps in the 3D case and 1.04$\times10^{7}$ steps in the 2D case corresponding to 48 hours and 4 hours, respectively (assuming a vacancy diffusion constant of $D_0$ = 3.6$\times10^{6}$ cm/s$^2$.\cite{Orwa2012}
In the 3D case, we assume a layer thickness of 2900 $\micro$m and 3.55 nm for the 2D case.
Once the vacancy centers the nitrogenated layer we form an NV with probability equal to the fractional density of nitrogen within the layer (10 ppm for the 3D layer and 2 ppm for the 2D layer).
Results of the calculation show that despite the 3D layer's increased nitrogen density and random walk (annealing) time, the 2D layer exhibits a $15\times$ enhancement in the ratio of NVs formed to initial P1 density.

We assume a homogeneous layer and so enforce periodic boundaries in the lateral direction.
Because we do not account for depletion of nitrogen upon NV formation, we expect the model to be valid in the low-vacancy limit but to break down when the density of NV centers becomes comparable to the remaining nitrogen.
Additionally, the model does not account for divacancy or vacancy cluster formation, nor does it consider Coulombic interactions (\textit{e.g.} V-P1 electrostatic attraction/repulsion).
A more complex model which takes into account nitrogen depletion and defect clustering would be needed for a more rigorous description of the N to NV conversion process.
The simulation code is given below:

\begin{lstlisting}
function [frac_conversion,frac_lost,frac_cut,Ndensdepth,NV_conversion] = 
vacancy_wander_2D3D_dens(Ndens,Ndepth,Nstraggle,Numruns,Nsteps,
vacancy_creation_depth)
%Ndens is volume density (3D)
%Nstraggle is the total width of the N layer.
%Ndepth is the center of the layer
numNV = 0;
numlost = 0;
numcut = 0;
Ndensdepth = zeros(Ndepth+ceil(Nstraggle),1);
for i = 1:Numruns
    
    zV = randi(vacancy_creation_depth); %Initialize the vacancy at the 
    bottom of the nitrogen layer
    j = 0; %Number of steps
    while j < Nsteps
        movedirection = randi(6);
        if movedirection == 1
            zV = zV - 1;
        elseif movedirection == 6
            zV = zV + 1;
        end
        
        if (zV > Ndepth - ceil(Nstraggle/2))&&(zV < Ndepth
        + ceil(Nstraggle/2))
        %If my vacancy is in the nitrogen layer
            testNV = rand; %generates [0,1)
            if testNV < Ndens %Ndens units of fractional carbons
                numNV = numNV+1;
                Ndensdepth(zV) = Ndensdepth(zV)+1;
                j = Nsteps+1; %breaking the loop
            end
        end
        
        if zV < 0
            numlost = numlost + 1;
            j = Nsteps+1; %breaking the loop
        end
        j = j + 1;
        if j == Nsteps
            numcut = numcut+1;
            j = Nsteps+1;
        end
    end 
    
    if mod(i,floor(Numruns/10)) == 0
        disp(strcat('Current progress =',num2str(floor(100*i/Numruns)),'%'))
    end
end

frac_conversion = numNV/Numruns;
frac_lost = numlost/Numruns;
frac_cut = numcut/Numruns;
NV_conversion = frac_conversion/(Ndens*Nstraggle);

end
\end{lstlisting}

\section{NV coherence}
\begin{figure}
    \centering
    \includegraphics{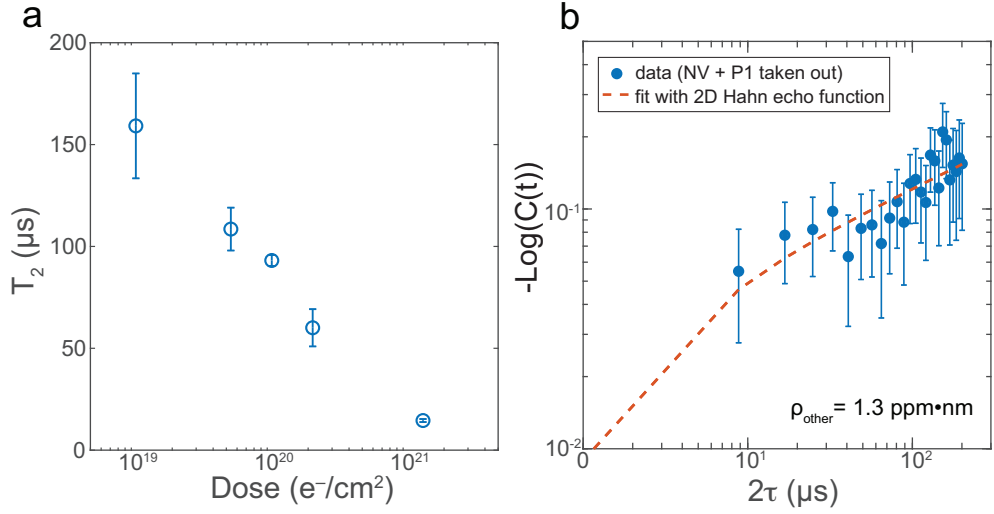}
    \caption{(a) T$_2$ as a function of electron irradiation dose measured in sample A. Error bars show standard error of the mean. (b) Log-log plot of the Hahn echo coherence decay, fit with a 2D model, yielding $\rho_{other}$ = 1.3 ppm$\cdot$nm}. 
    \label{fig:S5}
\end{figure}

To investigate the NV coherence in different defect environments, we surveyed the Hahn echo T$_2$ coherence time across all of the TEM irradiation spots on sample A.
Fig.~S\ref{fig:S5}a shows the T$_2$ as a function of electron irradiation dose, where the errors bars represent the standard error (four analyzed spots for the 10$^{19}$-10$^{20}$ dose range, three spots for 2$\times10^{20}$, and two spots for the 1$\times10^{21}$ dose).
The decrease in coherence with irradiation dose is partially explained by increasing NV density, but we hypothesize that a rising contribution to decoherence comes from an increasing population of other irradiation-induced defects.

Fig.~S\ref{fig:S5}b shows Hahn echo data taken on a region of TEM spot I on sample A where the NV and P1 contributions to decoherence ($\rho_{P1}$ = 6.0 ppm$\cdot$nm and $\rho_{NV}$ = 3.2 ppm$\cdot$nm) have been subtracted out from the data.
Re-fitting the Hahn echo reveals a remaining bath population $\rho_{other}$ of 1.3 ppm$\cdot$nm density.
This analysis shows that the majority of decoherence in our system comes from intentionally created defects (P1 and NV).
In other TEM irradiation spots, we observe varying densities of $\rho_{other}$ but the Hahn echo T$_2$ decays are still largely dominated by the intended spin-spin interactions.
Interestingly, the decoherence contribution from $\rho_{other}$ fits best with a 2D model across the TEM spots measured. 
This suggests that $\rho_{other}$ may consist of a nitrogen-related defect since vacancy creation occurs over a 50 $\micro$m depth with 200 keV irradiation.\cite{Kim2012}
Possible candidates for $\rho_{other}$ could be NVH or NVN, but more analysis beyond the scope of this work is needed to specifically identify these defect populations.

\section{Interpretation of decoherence decay signal}
In our analysis, we consider the probe NV center spin and surrounding noise bath (consisting predominantly of P1 and other NVs), coupled only by magnetic dipole-dipole interaction,
\begin{equation}
    H_z=\sum_i{\frac{Jf_i}{r_i^3}S^z_\text{NV}S^z_i},
\end{equation}
where coupling strength $J=\mu_0h\gamma^2/8\pi^2=2\pi\times52\text{ MHz·nm}^3$ for dipole-dipole interaction. 
$f_i$ captures the angular dependence of the interaction between the $i$-th spin in the noise bath and the probe NV. 
The probe NV axis is aligned with the external magnetic field and thus is also parallel to the quantized axis of P1 centers and other on-resonance NVs. 
We assume that the bath spins are not correlated with each other and the correlation function is simply $\langle 4S^z_i(t')S^z_i(t'')\rangle=\exp(-(t'-t'')/\tau_c)$, with $\tau_c$ being the correlation time. 
Here we also assume that the spin flip-flop noise of the bath demonstrates a Gaussian nature instead of telegraph behavior.\cite{Davis2021}

When considering an ensemble of NV centers, every NV has a different local spin bath environment. 
Assuming that the distribution of P1 centers around an NV is totally homogeneously random and the interaction between the bath spins is purely Ising, one can derive the experimental signal analytically via averaging over this positional randomness. 
The contribution from a specific group of spin-1/2 in the bath to the NV decoherence profile in a Hahn echo or DEER measurements is:
\begin{equation}
    C(t)=2\langle S^x_\text{NV}(t)\rangle 
    =\exp{[\rho\frac{D\pi^{(D-1)/2}\Gamma(-\frac{D}{2\alpha})\Gamma(-\frac{D}{2\alpha}+\frac{1}{2})\cos{(\frac{D\pi}{2\alpha})}}{\alpha\Gamma(\frac{D}{2}+1)2^{\frac{D}{\alpha}+1}}[\frac{|\bar{f}|J\chi(t)^{1/2}}{2}]^{\frac{D}{\alpha}}]}
\end{equation}
where $\rho$ is the D-dimensional density of that specific group of spins in the bath, and $|\bar{f}|=(\frac{\int |f|^{\frac{D}{\alpha}} d\Omega}{\int d\Omega})^{\frac{\alpha}{D}}$ is the averaged angular dependence over D-dimensional solid angle. 
$t$ is the time of the decoherence. 
$\chi(t)$ encodes the response of NV to the noise spectral density of that specific group of spins in the noise bath. 
Assuming ideal pulse sequences, if the pulse sequence's effective filter function applied on that group of spins-1/2 is a Hahn echo measurement, then $\chi(t)=2\tau_c t-2\tau_c^2(3+e^{-t/\tau_c}-4e^{-t/2\tau_c})$. 
For a Ramsey measurement, $\chi(t)=2\tau_c t-2\tau_c^2(1-e^{-t/\tau_c})$. 
For XY-8 measurement, we have $\chi(t)=t\tau_p^2/12\tau_c$, if $\tau_p\ll\tau_c$, where $\tau_p$ is the interval between the $\pi$ pulses in the sequence. %(For the density measurement sequence in Tim/Claire's paper, we get $\chi(t)=4\tau\tau_c-10\tau_c^2+2\tau_c^2e^{-2\tau/\tau_c}+(8e^{-\tau/\tau_c}-4e^{-2\tau/\tau_c})\tau_c^2e^{t/\tau_c}+4\tau_c^2e^{-t/\tau_c}$.) 
In a DEER measurement $\chi(t)$ for the group of spins that are driven by the $\pi$ pulse in the sequence is effectively the same as a Ramsey.

If the P1 centers are located in a 2D layer, after averaging over positional randomness, the signal of Hahn echo or DEER measurements can be written as:
\begin{equation}
    C(t)=\exp{[-\rho(aJ)^{2/3}\chi(t)^{1/3}]},
\end{equation}
where $a\approx2.626$ is a dimensionless constant. 
As for a 3D layer, the signal can be written as:
\begin{equation}
    C(t)=\exp{[-\rho aJ\chi(t)^{1/2}]},
\end{equation}
where for 3D bath $a\approx3.3178$ which is difference from when it is a 2D bath. 
For an effectively Hahn echo measurement, we denote the signal contribution from a specific group $\nu$ of spins in the noise bath as $E_\nu(t)$
\begin{equation}
    E_\nu(t)=\exp{[-\rho_\nu(aJ)^{2/3}[2\tau_c t-2\tau_c^2(3+e^{-t/\tau_c}-4e^{-t/2\tau_c})]^{1/3}]}.
\end{equation}
For an effectively Ramsey or DEER measurement, we denote the signal contribution from a specific group $\nu$ of spins in the noise bath as $R_\nu(t)$
\begin{equation}
    R_\nu(t)=\exp{[-\rho_\nu(aJ)^{2/3}[2\tau_c t-2\tau_c^2(1-e^{-t/\tau_c})]^{1/3}]}.
\end{equation}
For an effectively XY-8 measurement, we denote the signal contribution from a specific group $\nu$ of spins in the noise bath as $X_\nu(t)$
\begin{equation}
    X_\nu(t)\approx\exp{[-\rho_\nu(aJ)^{2/3}(\frac{\tau_p^2}{12\tau_c}t)^{1/3}]}.
\end{equation}

Now consider our systems where there are other on-resonance NVs, off-resonance NVs, two 1/8 groups of P1 centers and two 3/8 groups of P1 centers, so the signal of our Hahn echo measurement can be decomposed to:
\begin{equation}
    C_\text{Hahn}(t)=[E_{1/8}(t)]^2 [E_{3/8}(t)]^2 R_{\text{NV,on}}(t) E_{\text{NV,off}}(t).
\end{equation}
The signal of our DEER measurement, driving one of the 3/8 group, can be decomposed to
\begin{equation}
    C_\text{DEER}(t)=R_{3/8}(t) [E_{1/8}(t)]^2 E_{3/8}(t) R_{\text{NV,on}}(t) E_{\text{NV,off}}(t).
\end{equation}
Therefore, the DEER signal normalized by Hahn Echo should be:
\begin{equation}\label{eq:11}
    C(t)=\frac{R_{3/8}(t)}{E_{3/8}(t)}=\frac{\exp[-\rho_{3/8}(aJ)^{2/3}(2\tau_\text{c}t-2\tau_\text{c}^2(1-e^{-t/\tau_c}))^{1/3}]}{\exp[-\rho_{3/8}(aJ)^{2/3}(2\tau_\text{c}t-2\tau_\text{c}^2(3+e^{-t/\tau_c}-4e^{-t/2\tau_c}))^{1/3}]}
\end{equation}
where $\rho_{3/8}$ is the 2D density of the addressed P1 centers of 3/8 group. 
This signal should have a stretch power of $n\approx2/3$ when measuring on a time scale much shorter than correlation time, whereas if the spin bath is 3D it should be $n\approx1$ instead. 
In the case of either NV ionization or recombination present on a time scale of DEER and Hahn echo measurements, if performing differential measurements, the DEER signal normalized by Hahn echo signal should be immune to the varying of NV- state population $\rho_{\text{NV}^-}(t)$. 
We are mostly interested in the density of the spins so in order to simplify the fitting, we focus on the data where $t\ll\tau_c$. 
Under this assumption, we have
\begin{equation}\label{nvp1ramsey}
    C(t)\approx\exp[-\rho_{3/8}(aJ)^{2/3}t^{2/3}].
\end{equation}
Thus clearly, it has a stretch power of 2/3. 
Observation of the stretch power can indicate the nature of dimensionality of the spin bath.

We now apply the analysis to XY-8 measurement in order to extract the NV density. 
The signal of the XY-8 is composed of the contribution from NVs that have same axis via a Ramsey filter function and the contribution from other spins in the bath via a XY-8 filter function.
\begin{equation}
    C_\text{XY8}(t)=R_{\text{NV,on}}(t)[X_{1/8}(t)]^2 [X_{3/8}(t)]^2 X_{\text{NV,off}}(t).
\end{equation}
We expect the contribution from on-resonance NVs is dominant over P1 centers. 
Note that even though the NVs are spin-1 instead of spin-1/2, the Ramsey signal $R(t)$ is still equivalent to having a spin-1/2 bath. The effective coupling strength $J_\text{eff}$ for NV-NV intra-group interaction here is equal to $2J=4\pi\times\SI{52}{MHz\cdot nm^3}$. 
In order to better extract the density, we measure by fixing $\tau_p$ at a very small value and vary the number of XY-8 pulses. 
Thus we can ignore the contribution from P1 centers and other off-resonance NVs, and we approximately have
\begin{equation}
    C_\text{XY8}(t)\approx\exp[-\rho_\text{NV,on}(aJ_\text{eff})^{2/3}(2\tau_\text{c}t-2\tau_\text{c}^2(1-e^{-t/\tau_c}))^{1/3}].
\end{equation}
If we also assume $t\ll\tau_{c}$, we have
\begin{equation}\label{nvnvramsey}
    C_\text{XY8}(t)\approx\exp[-\rho_\text{NV,on}(aJ_\text{eff})^{2/3}t^{2/3}].
\end{equation}

The results in equation (\ref{nvnvramsey}) and (\ref{nvp1ramsey}) can also be verified with a numerical simulation, assuming the bath spins do not change spin state during one shot of the pulse sequence measurement under the approximation of long correlation time. 
The simulation is generated as follows: we consider a central probe NV interacting with a bath of other P1 centers or NVs, placed randomly in a thin slab of thickness of 0.1 nm and a certain spin density. 
Then we select a random spin configuration for the bath spins $S_i^z$, and assume not involving any spin flip-flops. 
If considering P1 centers, the $S_i^z$ has 50\% probability to be +1/2 and 50\% probability to be -1/2. 
We assume the quantization axis of P1s has equal probability to be in either one of the four Jahn-Teller axes (same as crystal lattice axes). 
If considering NVs bath, then it is 0 or -1, assuming we are driving the $\ket{0}\ket{-1}$ transition. 
Since we are considering XY8 measurement which is detecting only one group of NVs, so the quantization axis of NVs in the bath is the same axis as the probe NV. 
We compute the Ramsey signal for the probe NV and average over many such samples. 
The resulting signal curve then is then compared to the theory and shows agreement.

Similarlly, one could also derive the signal from the NV-NV DEER measurement and verified by the numerical simulation. 
For example, for a NV-NV DEER XY8 sequence, when the three other NV groups cannot be individually driven with appropriate $\pi$ pulse duration, we have
\begin{equation}
    C_\text{XY8DEER}(t)> R_{\text{NV,total}}(t)[X_{1/8}(t)]^2 [X_{3/8}(t)]^2.
\end{equation}
The nature of the NV-NV DEER measurement is similar to XY8. 
As mentioned in previous section, this measurement is underestimating the total density of NV, because the Rabi frequencies of difference NV groups are different but we drive three NV groups together with the same $\pi$ pulse duration in the sequence.

\bibliography{2D_si} 
\bibliographystyle{ieeetr}